# Queue Management in Network Processors

I. Papaefstathiou[1], T. Orphanoudakis[2], G. Kornaros[2], C. Kachris[2], I. Mavroidis[2], A. Nikologiannis[2]

| 1 Foundation of Research & Technology Hellas (FORTH), Institute of Computer Science (ICS), Vassilika Vouton, GR71110, Heraklio, Crete, Greece | 2 Ellemedia Technologies 223, Siggrou Av, GR17121, Athens, Greece {fanis,kornaros,kachris,jacob,anikol}@ellemedia.com |

ygp@ics.forth.gr

*Abstract:* - One of the main bottlenecks when designing a network processing system is very often its memory subsystem. This is mainly due to the state-of-the-art network links operating at very high speeds and to the fact that in order to support advanced Quality of Service (QoS), a large number of independent queues is desirable. In this paper we analyze the performance bottlenecks of various data memory managers integrated in typical Network Processing Units (NPUs). We expose the performance limitations of software implementations utilizing the RISC processing cores typically found in most NPU architectures and we identify the requirements for hardware assisted memory management in order to achieve wire-speed operation at gigabit per second rates. Furthermore, we describe the architecture and performance of a hardware memory manager that fulfills those requirements. This memory manager, although it is implemented in a reconfigurable technology, it can provide up to 6.2Gbps of aggregate throughput, while handling 32K independent queues.

## 1   Introduction

To meet the demand for higher performance, flexibility, and economy in emerging multi-service broadband networking systems, an alternative to Application Specific Integrated Systems (ASICs), which have been traditionally used to implement packet-processing functions in hardware, the so called Network Processors or Network Processing Units (NPUs), has emerged. NPUs can be broadly defined as System-on-Chip (SoC) architectures integrating multiple simple processing cores (so as to exploit parallelism and/or pipelining in order to increase the supported network throughput) and performing complex protocol processing at multi Gigabit per second rate. These processing cores are either Reduced Instruction Set Computing (RISC) CPUs, or dedicated hardware engines for specific complex packet processing functions that require wire-speed performance like classification, per-flow queuing, buffer and traffic management.

Most modern networking technologies (like IP, ATM, MPLS etc.) share the notion of connections or flows (we adopt the term "flow" hereafter), that represent data transactions in specific time periods and between specific end-points in the network. Depending on the applications and algorithms used, the network processor typically has to manage thousands of flows, implemented as packet queues in the processor packet buffer [1]. Therefore, effective queue management is a key to high-performance network processing as well as to reduced development complexity. The focus of this paper is twofold: first we quantify the bottlenecks of employing packet queues in legacy general purpose processing units; then we briefly present an FPGA-based queue management system, which can scale efficiently and provide an efficient solution for demanding applications. We claim that this hardware module is a very useful component for every networking system that manipulates queues since: a) it supports a large number of simple request-acknowledge interfaces, b) it executes a large number of general instructions and c) it can handle either fixed size or variable length pieces of data. In particular, we believe that this system will be a valuable add-in, likewise a co-processor in a separate FPGA, for the commercial ASIC NPs that have no dedicated memory handling hardware.

In order to accurately evaluate the effectiveness of the various software and hardware schemes, we first briefly describe, in Section 2, a number of existing NPU architectures focusing on their memory management optimizations and then we analyze the necessary external memory bandwidth needed for implementing a general queue management system in such an NPU. In particular, in section 3 we analyze the performance bottlenecks of a reference such system, examining the accesses to external memories in isolation, based on the memory access patterns of real-world network applications. In section 4 we present an analysis regarding the performance of a queue management implementation on a widely used Network Processor and in section 5 we proceed in a more detailed analysis expanding our results to a generic NPU prototype architecture. After summarizing our experiences from software-based implementations in section 5.4, in section 6, we present our FPGA-based queue management system. The conclusions of our paper are finally outlined in section 7.

## 2   Related Work: Memory Management in Network Processors

The main driver for sophisticated memory management systems, in almost ever NPU, is the requirement for data packets to be stored in appropriate queue structures, either before or after processing, and then to be selectively transmitted. These queues of packets should not be, in the majority of cases, organized as simple FIFOs, but instead should provide the means to access certain parts of their structures (i.e. access packets which reside in a specific position in the queue e.g. head or tail of the queue etc.). In order to efficiently cope with these requirements several solutions, based on dedicated hardware modules, have been proposed. Initially those modules were targeting high-speed ATM networks, where, due to the fixed ATM cell size, very efficient queue management was possible ([2], [3]), while later on they have been extended to the management of queues of variable-size packets [4]. The basic advantage of these hardware implementations is, obviously, the high throughput they can achieve. On the other hand the functions they can provide (e.g. single vs. double linked lists, operations in the head/tail of the queue, copy operations etc.) need to be selected very carefully, when initially designing the hardware module, in order for those systems to be efficient for the majority, at least, of the network applications. Several trade-offs between dedicated hardware modules and implementations in software, for ATM networks, have been exposed in [5].

In general, several commercial NPUs follow a hybrid approach for efficient memory management: they utilize specialized hardware units that implement certain memory access sub-operations, but they do not provide a complete queue management hardware implementation. The first generation of the Intel NPU family, the IXP1200 [6], provides an enhanced SDRAM control unit, which supports single byte, word, and long-word write capabilities using a read-modify-write technique and may reorder SDRAM accesses for best performance (the benefits of this feature will also be exposed in the following section). The SRAM Control Unit of the IXP1200 also includes an 8-entry Push/Pop register list for fast queue operations. Although these hardware enhancements improve the performance of typical queue management algorithms they cannot keep up with the requirements of high-speed networks. Therefore the next generation IXP-2400 provides high-performance queue management hardware that efficient supports the enqueue and dequeue operations [6]. Following the same approach the PowerNP NP4GS3 incorporates dedicated hardware acceleration for cell enqueue/dequeue operations in order to manage packet queues [7]. Freescale's C-5 NPU also provided memory management acceleration hardware [8], which is probably not adequate, though, to cope with demanding applications that require frequent access to packet queues. Therefore, the same company has also manufactured the Q-5 Traffic Management Coprocessor, which consists of dedicated hardware modules designed to support traffic management for up to 128K queues at a rate of 2,5 Gbps [9].

## 3 External DRAM Memory Bottlenecks

Since a DRAM offers high throughput and very high capacity per unit cost, packet buffers are stored in external DRAMs in most of today's NPUs. Among DRAM technologies, we focus our analysis on DDR-SDRAM because it achieves very high performance while it is very cost-effective due to its widespread use.

The DDR technology provides 12.8 Gbps of peak throughput when using a 64-bit data bus at 100 MHz with double clocking (i.e. 200 Mb/sec/pin). A DIMM module provides up to 2 GB of total capacity and it is organized into 4 or 8 banks in order to provide interleaving (i.e. to allow multiple parallel accesses). However, due to the bank-precharging period (i.e. when the bank is busy), successive accesses[1] to the same bank may be performed every 160 ns. When a memory transaction tries to access a currently busy bank we say that a bank conflict has occurred. This conflict causes the new transaction to be delayed until the bank becomes available, thus reducing memory utilization. In addition, interleaved read and write accesses also reduce the mean memory utilization because they have different access delays[2]. By simulating a behavioral model of a DDR-SDRAM memory, we have estimated the impact of bank conflicts and read-write interleaving on memory utilization. Random bank access patterns were simulated as a realistic common case for typical network applications incorporating a large number of simultaneously active queues. The results of this simulation, for a range of banks, are presented in the two left columns of Table 1.

| banks | No Optimization Throughput Loss | | Optimization Throughput Loss | |
|---|---|---|---|---|
| | Bank conflicts | Bank conflicts + write-read interleaving | Bank conflicts | Bank conflicts + write-read interleaving |
| 1 | 0.750 | 0.75 | 0.750 | 0.750 |
| 4 | 0.522 | 0.5 | 0.260 | 0.331 |
| 8 | 0.384 | 0.39 | 0.046 | 0.199 |
| 12 | 0.305 | 0.347 | 0.012 | 0.159 |
| 16 | 0.253 | 0.317 | 0.003 | 0.139 |

Table 1: DDR-DRAM throughput loss using 1 to 16 banks

We considered aggregate accesses from 2 write and 2 read ports[3]. By serializing the accesses from the 4 ports in a round-robin manner we measured the throughput loss presented in Table 1. However, if the accesses of the 4 ports are scheduled in a more efficient way, we can achieve a lower throughput loss by reducing bank conflicts. The simplest approach is to effectively reorder the accesses of the 4 ports, in order to minimize bank conflicts. This can be performed by organizing pending accesses into 4 FIFOs (1 FIFO per port). In every access cycle the scheduler checks the pending accesses from the 4 ports for conflicts and selects an access that addresses a non-busy bank. The information for bank availability is achieved by keeping the memory access history (it remembers the last 3 accesses). In case that more than one accesses are eligible (i.e. belong to a non-busy bank), the scheduler selects one of the eligible accesses in a round-robin order. In case that no pending access is eligible, the scheduler sends a no-operation to the memory, losing an access cycle. The results of this simple optimization are presented in Table 1. Assuming 8 banks per device, this very simple optimization scheme reduces the throughput loss by 50% in comparison with the not-optimized one.

---

[1] A new read/write access to 64-byte data blocks can be inserted to DDR-DRAM every 4-clock-cycles (access cycle = 40 ns).

[2] Write access delay = 40 ns, Read access delay = 60 ns. When write accesses occur after read accesses, the write access must be delayed 1 access cycle.

[3] A write and a read port from/to the network, a write and a read port from/to an internal processing unit.

## 4  Queue Management on the IXP1200

As it was described in Section 2, the most straightforward implementation of memory management in NPUs is based on software executed by one or more on-chip microprocessors. Apart from the memory bandwidth, which we examined in isolation in the previous section, a significant factor that affects the overall performance of a queue management implementation is the combination of the processing and data transfer latency (including the latency for the communication with the external memories and their controllers). Additionally, since dynamic memory management is usually based on the implementation of linked list structures, the respective pointer storage is almost always performed on SRAM memories; this is due to the fact that the pointer manipulation tasks need short accesses compared to the burst data accesses needed for buffering network packets. The very frequent pointer manipulation functions can also be the bottleneck of the queue management system depending on the application requirements and the hardware architecture. Therefore, the overall actual performance of a memory management scheme can only be accurately evaluated at the system level. We used Intel's IXP1200 as a typical NPU architecture and we provide indicative results regarding the maximum throughput that can be achieved when implementing queue management on this NPU.

The IXP1200 consists of 6 simple RISC processing microengines [6] running at 200MHz. When porting the queue management software to those RISC-engines, special care should be given so as to take advantage of the local cache memory (called "Scratch memory") as much as possible. This is because any accesses to the external memories take a very large number of clock cycles. One can argue that using the multithreading capability of the IXP, someone can hide this memory latency. However, as it was demonstrated in [10], the overhead for the context switch, in the case of multithreading, exceeds the memory latency and thus this IXP feature cannot increase the performance of the memory management system, when external memories should be accessed.

Even by using a very small number of queues (i.e. less than 16), so as to be able to keep every piece of control information in the local cache and in the IXP's registers, we have measured that each microengine cannot service more than 1 Million Packets per Second (Mpps). Or, in other words, the whole of the IXP cannot process more than 6Mpps. Moreover, if 128 queues are needed, and thus some external memory accesses are necessary, each microengine can process at most 400Kpps. Finally, for 1K queues the peak bandwidth that can be serviced by all 6 IXP microengines is about 300Kpps, which agrees with the result in [11]. Since, in the worst case, the Ethernet packets are 64-byte long, we claim that the whole of the IXP cannot support more than 150Mbps of network bandwidth, even if only 1K queues are needed. We summarize the above throughput results in Table 2.

From the above it can easily be derived that this software approach, cannot cope with today's state-of-the-art network links, if the network application involves the handling of more than a hundred separate queues.

| Num of Queues | 1 Microengine | 6 Microengines |
|---|---|---|
| 16 | 956 Kpps | 5.6 Mpps |
| 128 | 390 Kpps | 2.3 Mpps |
| 1024 | 60 Kpps | 0.3 Mpps |

Table 2: Maximum Rate Serviced when queue management runs on IXP 1200

## 5  Custom Software Implementation of Memory Management on a Generic NPU

In order to be able to experiment with different design alternatives and perform detailed measurements, we have implemented ourselves, a typical reference NPU. With the aid of a state-of-the-art FPGA that provides hard macros of very sophisticated embedded RISC cores, we have implemented the core design of an NPU. The architecture of the system, which was ported to a Xilinx Virtex-II Pro device [14], is depicted in Figure 1. As it is shown, the 64-bit Processor local bus (PLB) is used as the system bus, at a clock frequency of 100MHz. The PowerPC 405 is used as the main processor. The OCM Controller is used to connect the PowerPC with the Specialized Instruction and Data Memory (16KBytes each). The size of the code used for memory management is small enough to fit in this small instruction memory. The packets are stored in an external DDR DRAM using a sophisticated DDR controller, while the queue information (mainly pointers) is stored in an external ZBT SRAM, using Xilinx's PLM External Memory Controller (EMC). In order to measure the performance of the system when real network traffic is applied to it, an Ethernet MAC port has been used. The MAC Core (provided by OpenCores) uses two WishBone(WB) Compatible ports. The first port is attached to the PLB Bus, through the PLB-to-WB Bridge, and is used for control. The second port is attached to a 4 Kbytes Dual Port internal Block RAM (DP-BRAM), and is used to store temporarily the in-coming and out-going Ethernet packets. With the aid of this on-chip DP-BRAM data transfers between the network interface and the queue manager (i.e. processor and buffer memory) can be achieved very efficiently.

### 5.1.  Configuration

The PowerPC has been configured to use the instruction and data cache, both in write back mode. The PowerPC and PLB Bus clock frequency has been set to 100MHz and the DDR controller is configured in burst mode. Finally, the code has been compiled using GCC optimization level 2 and then handcrafted. The frequency selection was dictated by the implementation timing requirements. A state of the art embedded RISC core though, (like the PowerPC core provided in the Xilinx Virtex II Pro family) can easily reach operation frequencies in the range 200-300 MHz (even in this reference FPGA device), so we also compare the performance in those projected range of frequencies. Note that the design of Figure 1 represents a typical organization of an NPU core design, where the PowerPC is used as a typical on-chip embedded processor; the PowerPC is may even be more powerful when compared to the typical such cores used in commercial NPs.

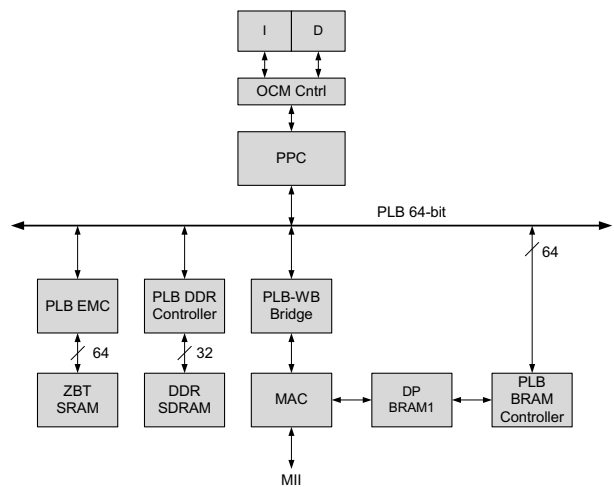

Figure 1: NPU core architecture set-up on the Xilinx Virtex-II Pro FPGA platform

### 5.2. Queue structure

We implemented queues of packets as single-linked lists. The incoming data items are partitioned into fixed size segments of 64 bytes each. Our implementation organizes the incoming packets into queues and handles and updates the data structures kept in the pointer memory. A free-list keeps the free parts of the memory, at any given time, and a queue-table contains the header of all the employed queues.

The Queue Manager supports mainly the following functions:
- Enqueue Segment
- Dequeue Segment
- Enqueue Free List
- Dequeue Free List

Each segment function is analyzed into separate segment and free list sub-operations. For example, the enqueue packet operation is analyzed into the following steps: First a new pointer is allocated from the free list, then this pointer is stored to the queue list and then the data are transferred to the memory.

### 5.3. Performance evaluation

Table 3 shows the number of cycles for the execution of each segment operation. For a 100Mbps network and a minimum packet length of 64 bytes the available time to serve this packet is 5.12 μsec.

| Function | Cycles | |
|---|---|---|
| | **Enqueue** | **Dequeue** |
| Dequeue Free List | 34 | 42 |
| Enqueue Segment | 46/68[*] | 52 |
| Copy a segment | 136 | 136 |
| *Total* | *216/238* | *230* |

[*]46 for the first segment of the packet, 68 for the rest

Table 3: Cycles per packet operation

Let assume that the PowerPC's clock frequency is set to 100 MHz, then the available time for processing a single packet is 512 clock cycles for a half-duplex network, or 256 cycles for a full duplex network. This means that for the queue management only, all the available processing capacity of the PowerPC core has to be used so as to support a full duplex 100Mbps line. In other words, the PowerPC cannot afford to further manipulate the packet, thus another processor must be used for further processing. The majority of the cycles are spent waiting for the data from the memory and for the transactions over the PLB bus. Even if the processor operation frequency is set to 400MHz, the improvement in the overall performance would not be significant, since the maximum frequency of the PLB bus, in the state-of-the-art reconfigurable chip is 200MHz and in general it is hard to clock a bus, such as the PLB, at more than 200MHz even in an ASIC.

As Table 3 demonstrates, half of the cycles are used to copy the data of the segment. A major improvement is to exploit the "line transactions" of the PLB. In this case, the PowerPC execute the line transactions over the bus using the data cache unit as a temporary buffer [12]. Using this configuration a segment can be retrieved from the BRAM and stored into the data cache in only 12 cycles (9 cycles for 9 double words and 3 cycle latency). Thus, the total number of cycles to copy a segment becomes:

$$T_C = (T_R + T_l) + (T_W + T_l) = 2*(9+3)=24 \text{ cycles}$$

where $T_R$ denotes the number of cycles to read a segment from the on-chip buffer (Xilinx BRAM block), $T_W$ denotes the number of cycles to write a segment to the DDR DRAM and $T_l$ denotes the 3-cycles bus latency. Thus, the total number of cycles to enqueue and dequeue a packet becomes 128 and 118 respectively, which dictates that the 100MHz PowerPC would sustain up to about 200 Mbps throughput.

Another improvement would be to use a sophisticated DMA controller like the one in [13]. In this case, four 32-bit registers (DMA control, source/destination address and length registers) have to be set before each transaction [14]. However, each single PLB write transaction needs 4 cycles, thus we need at least 16 cycles to initiate the DMA transfer and at least 34 cycles to copy the data from the BRAM to the DRAM or vice versa. Note that the total time per operation is approximately the same as before. Hence, the overall throughput does not increase significantly, but in this configuration the processor has additional available processing power for other applications, due to the offloading of the data copying tasks to the DMA engine.

## 5.4 Impact on system level design

The results of the previous section provide some insight on the limitations of software-based queue management implementations. These results can be roughly summarized in the following "rule-of-thumb": the clock frequency of the system is proportional to the network bandwidth supported, since a system with a 100MHz microprocessor seems to be adequate to handle only a full duplex 100Mbps network link. Of course, the supported throughput can be increased by employing enhanced memory transfer techniques, more efficient buses etc. In any case, the performance limitations of the software approach, probably, make it unsuitable for Gigabit networks. The trade-off is throughput vs. programmability. In case the NPU architecture targets a wide variety of applications with moderate throughput requirements (e.g. low-end wireline or wireless LANs, access or edge network equipment) and the application requirements may change over time, the inherent programmability of the embedded multiprocessor architectures offers an adequate solution. However, more demanding applications in terms of target link rates or amount of packet operations and queue manipulations may easily consume all the available processing resources even when advanced VLSI technologies are employed (in which case the final end-system cost becomes an issue, since additional processing power will not come for free even when technology makes it feasible). Efficient application-specific hardware engines seem to be the only solution in this case.

## 6 An FPGA-based Memory Management System (MMS)

The hardware-oriented approach addresses the limitations identified in the previous sections. In order to achieve efficient memory management, in hardware, the incoming packets are partitioned into fixed size segments of 64 bytes each. The segmented packets are stored in the data memory, which is segment aligned. The MMS performs per flow queuing for up to 32 K flows; each packet is assigned to a certain flow. The MMS offers a set of operations on the segmented packet, for flexible queue management, such as:
1. Enqueue one segment
2. Delete one segment or a full packet
3. Overwrite a segment
4. Append a segment at the head or tail of a packet
5. Move a packet to a new queue

These functions facilitate the execution of the basic packet forwarding operations; for instance segmentation & reassembly, protocol encapsulation, header modification. By supporting those operations, as shown in [4], we have managed to accelerate several real world network applications such as:
- Ethernet switching (with QoS e.g. 802.1p,802.1q)
- ATM switching
- IP over ATM internetworking
- IP routing
- Network Address Translation
- PPP (and others) encapsulation

The MMS uses a DDR-DRAM for data storage and a ZBT SRAM for segment and packet pointers. Thus, all manipulations on data structures (pointers) occur in parallel with data transfers, keeping DRAM accesses to a minimum. Figure 2 shows the architecture of the MMS. It consists of five main blocks: Data Queue Manager (DQM), Data Memory Controller (DMC), Internal Scheduler, Segmentation Block and Reassembly Block.

These blocks operate in parallel in order to increase performance. The internal scheduler forwards the incoming commands from the various ports to the DQM giving different service priorities to each port. The DQM organizes the incoming packets into queues. It handles and updates the data structures kept in the Pointer memory. The DMC performs the low level read and write segment commands to the data memory; it issues interleaved commands so as to minimize bank conflicts. The In/Out and the CPU interfaces can be connected to numerous physical network interfaces and to a large number of CPU cores.

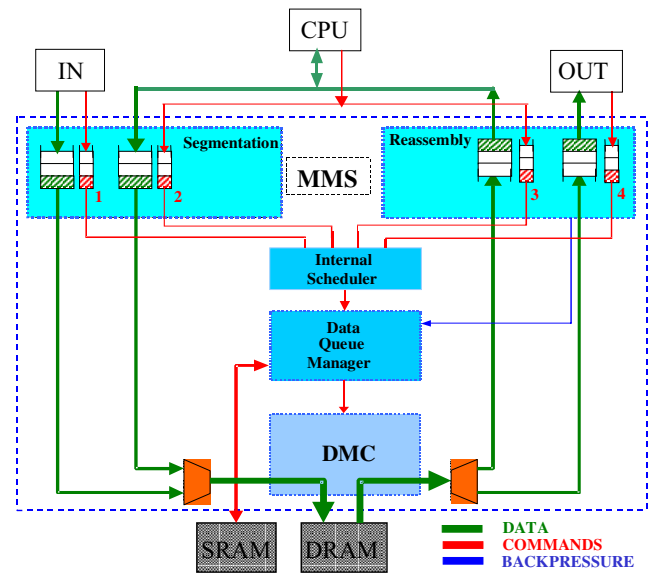

Figure 2: MMS Architecture

The MMS is a generic queue management block that can be easily incorporated in any networking embedded system that can handle queues.

## 6.1 Experimental results

Extensive experiments of the MMS were performed, in the framework described in the last section, and by using micro-code specifically developed for the embedded CPUs of the reference hardware platform.
Table 4 shows the measured latency of the segment commands. The actual data accesses at the Data Memory can be done, almost, in parallel with the pointer handling. In particular, a data access can start right after the first pointer memory access of each command has been completed. This is achieved because the pointer memory accesses, of each command, have been scheduled in such as way that the first one provides the corresponding Data memory address.

| Simple Commands | Clock Cycles |
|---|---|
| Enqueue | 10 |
| Read | 10 |
| Overwrite | 10 |
| Move_ | 11 |
| Delete | 7 |
| Overwrite_Segment_length | 7 |
| Dequeue | 11 |
| Overwrite_Segment_length&Move | 12 |
| Overwrite_Segment&Move | 12 |

Table 4: Latency of the MMS commands

The MMS latency has been measured for a system that has a conservative clock of 125 MHz (according to the synthesis and placement and routing tools, the MMS can work at more than 200MHz in a 0.18μm CMOS technology). Table 5 shows the MMS average latency for different loads. The total latency of a command consists of three parts: the FIFO delay, the execution latency and the data latency. MMS keeps incoming commands in FIFOs (one per port) so as to smooth the bursts of commands that may arrive simultaneously at this module. The latency that a command suffers, until it reaches the head of the FIFO, is the FIFO latency. As soon as a command reaches the head of the FIFO it starts its execution by accessing the pointer memory. The latency introduced from this point until the execution is completed is the execution latency. This latency defines the time interval between two successive commands; in other words it states the MMS processing rate. Finally, the delay required to read or write a data segment along including the possible DRAM bank conflict delay is called the data latency. Since the MMS accesses the pointer memory in parallel with the DRAM, the execution accounts only for 10.5 cycles of overhead delay. The MMS can handle one operation per 84 ns or 12 Mops/sec operating at 125MHz. In other words, and since each operation is executed on 64-byte segments, the overall bandwidth the MMS supports is 6.145Gbps.

| Overall Load (Gbps) | FIFO delay | Execution delay | Data delay | Total delay per command |
|---|---|---|---|---|
| | clock cycles | | | |
| 6.14 | 68 | 10.5 | 31.3 | 109.8 |
| 4.8 | 57 | 10.5 | 30.8 | 98.3 |
| 4 | 20 | 10.5 | 30 | 60.5 |
| 3.2 | 20 | 10.5 | 29.1 | 59.6 |
| 1.6 | 20 | 10.5 | 28 | 58.5 |

Table 5: MMS Delays

## 7 Conclusions

It is widely supported that one of the most challenging tasks in network processing is the memory/queue handling. In this paper, we have first analyzed the memory access characteristics and the processing requirements of some common queue manipulation functions. Our performance evaluation of queue management implementations, both on the commercial IXP1200 NPU, as well as on our reference prototype architecture, exposes the significant processing resources required when general-purpose RISC engines are used to implement queue manipulation functions. Those results show that even with state-of-the-art VLSI technology and processor frequencies in the order of several hundreds MHz, a single processor can only achieve a throughput in the order of hundreds of Mbps (and for a moderate number of queues). Hence, we claim that, in order to support the multi Gigabit per second rates of today's networks we need specialized hardware modules. In this paper we also briefly presented such a hardware module, which supports up to 6.2Gbps of network bandwidth when implemented on an FPGA. Since the hardware cost of the device is limited, we claim that such a hardware subsystem significantly increases the overall network processing performance at an acceptable cost.

*Acknowledgments*
This work was performed in the framework of the WEBSoC project, which is partially funded by the Greek Secretariat of Research & Technology.

**References**
[1] V. Kumar, T. Lakshman, and D. Stiliadis, "Beyond best-effort: Router architectures for the differentiated services of tomorrow's internet," IEEE Communications Magazine, pp. 152-164, May 1998.
[2] J. Rexford, F. Bonomi, A. Greenberg, A. Wong, "Scalable Architectures for Integrated Traffic Shaping and Link Scheduling in High-Speed ATM Switches," IEEE Journal on Selected Areas in Communications, pp 938-950, June 1997.
[3] D. Whelihan and H. Schmit, "Memory optimization in single chip network switch fabric", Proceedings of the 39th DAC, New Orleans, Louisiana, USA, June 2002.
[4] G. Kornaros, I. Papaefstathiou, A. Nikologiannis, N. Zervos "A Fully-Programmable Memory Management System Supporting Queue Handling at Multi Gigabit rates", IEEE, ACM, Proceedings of the 40th DAC, Anaheim, California, U.S.A., June 2-6, 2003.
[5] D. N. Serpanos, P. Karakonstantis, "Efficient Memory Management for High-Speed ATM Systems", Design Automation for Embedded Systems, Kluwer Academic Publishers, 6, 207–235, April 2001.
[6] S. Lakshmanamurthy, et. al. "Network Processor Performance Analysis Methodology", Intel Technology Journal Vol. 6 Issue 3, 2002.
[7] James Allen, et. al. "PowerNP network processor: hardware, SW and applications," IBM Journal of Research and Development, vol. 47, no. 2-3, pp. 177-194, 2003.
[8] C-port Corp., "C-5 Network Processor Architecture Guide", C5NPD0-AG/D, May 2001.
[9] Motorola Inc., "Q-5 Traffic Management Coprocessor Product Brief", Q5TMC-PB, December 2003
[10] W. Zhou et. al., "Queue Management for QoS Provision Build on Network Processor" 9th IEEE Work-shop on Future Trends of Distributed Computing Systems (FTDCS'03), May 28 - 30, 2003 San Juan, Puerto Rico
[11] Tammo Spalink, Scott Karlin, Larry Peterson, Yitzchak Gottlieb, "Building a Robust Software-Based Router Using Network Processors", 18th ACM Symposium on Operating Systems Principles (SOSP'01), Chateau Lake Louise, Banff, Alberta, Canada, October 2001.
[12] PowerPC 405 Processor Block Reference Guide, September 2, 2003, Xilinx Inc.
[13] Meleis, H.E., Serpanos, D.N., "Designing communication subsystems for high-speed networks", IEEE Network, Vol. 6, Issue 4.
[14] Direct Memory Access and Scatter Gather Datasheet, Version 2.2, January 2003, Xilinx Inc.